\documentclass[aps, preprint, eqsecnum, showpacs, epsfig]{revtex4}
\usepackage{graphics} 
\def \PY {Percus-Yevick}
\def \GB {Gay-Berne}
\def \tm {temperature}
\def \be {\begin{equation}}
\def \ee {\end{equation}}
\def \ba {\begin{eqnarray}}
\def \ea {\end{eqnarray}}
\def \bm {\begin{displaymath}}
\def \em {\end{displaymath}}
\def \br {{\bf r}}
\def \bom {{\bf \Omega}}
\def \bn {{\bf 1}}
\def \bp {{\bf 2}}
\def \bs {{\bf 3}}
\def \pcf {pair correlation functions}
\def \dft {density functional theory}
\begin{document}
\title{The Structure and Freezing of fluids interacting via the
Gay-Berne (n-6) potentials}
\author{ Ram C. Singh$^\dagger$, Jokhan Ram and Yashwant Singh }
\affiliation{ Department of Physics, Banaras Hindu University, \\
          Varanasi 221 005, India\\ 
$\dagger$ Department of Physics, Meerut Institute of Engineering
          and Technology, Meerut-250 002,
          India}
\date{\today}
\begin{abstract}

We have calculated the pair correlation functions of a fluid interacting
via the Gay-Berne(n-6) pair potentials using the \PY\ integral equation
theory and have shown how these correlations depend on the value of n 
which measures the sharpness of the repulsive core of the pair potential.
These results have been used in the density-functional theory to locate
the freezing transitions of these fluids. We have used two different versions
of the theory known as the second-order and the modified weighted
density-functional theory and examined the freezing of 
these fluids for $8 \leq n \leq 30$
and in the reduced temperature range lying between 0.65 and 1.25 into
the nematic and the smectic A phases. For none of these cases smectic A 
phase was found to be stabilized though in some range of temperature
for a given $n$ it appeared as a metastable state. We have examined the 
variation of freezing parameters for the isotropic-nematic transition with
temperature and $n$. We have also compared our results with simulation results
wherever they are available. While we find that the density-functional
theory is good to study the freezing transitions in such fluids the 
structural parameters found from the \PY\ theory need to be improved 
particularly at high temperatures and lower values of $n$.
\end{abstract}                                                               
\maketitle
\section{Introduction }

In the case of non-spherical molecules the anisotropic nature of the 
intermolecular interactions can give rise to new phases (liquid crystals)
\cite{1} that are absent when simple spherical molecules are considered.
Depending upon the shape and the size of molecules and upon the external
parameters (temperature, pressures, etc.) a system may show a wide variety
of phenomena and transitions in between the isotropic liquid
and the crystalline solid. All these phases including that of the
isotropic liquid and the crystalline solids are characterized by the
average positions and orientations of molecules and by the intermolecular
spatial and orientational correlations. The determination of phase diagram
of such a system  from the intermolecular potential is one of the most
challenging problems of the statistical mechanics.

The molecules of systems  which exhibit liquid crystalline phases are 
generally large and have group of atoms with their own local features.
In general it is difficult to know the true nature of the potential
energy of interaction between such molecules. Attempts have, however, been made
to find the potential energy of interactions between two such molecules
using different approximations. One such method is to sum the interatomic
or site-site potentials between atoms or between interaction sites. In 
another and more convenient approach one uses rigid molecules approximation
in which it is assumed that the intermolecular potential energy depends
only on the position of the centre of mass and on their orientations.
If, however, our interest is to relate the phases formed and their
properties to the essential molecular factor responsible for the existence 
of liquid crystals, it is desirable to use a phenomenological description,
either as a straightforward model unrelated to any particular physical
systems or as  a basis for describing by means of adjustable parameters
between two molecules. Most commonly used models are hard-ellipsoids 
of revolution, hard spherocylinders \cite{2}, cut-sphere, the Kihara
core model \cite{3} and the \GB\ \cite{4} model. All these are single site
models and refer to rigid molecules of cylindrical symmetry. Even for
these simple models calculating the complete phase diagram is difficult.
 
 The Gay-Berne potential, in particular, is proving to be a valuable
model with which to investigate the behavior of liquid-crystals
in recent years using computer simulation techniques \cite{5,6,7}.
In this paper we consider a general Gay-Berne (GB) model with n-6
dependence on the shifted and scaled separation, R, between the 
uniaxial particles.

\be\
u({\hat {\bf e_i}}, {\hat {\bf e_j}}, {\hat{\bf r}}) = 
4 \epsilon ({\hat {\bf e_i}}, {\hat {\bf e_j}}, {\hat{\bf r}})
(R^{-n}-R^{-6})
\ee\
where 
\begin{equation}
R = \frac{(r-\sigma({\hat{\bf e_i}}, {\hat{\bf e_j}}, {\hat{\bf r}}) + 
\sigma_0)}{\sigma_0}
\end{equation} 

While unit vectors ${\hat{\bf e_i}},{\hat{\bf e_j}}$ indicate the 
orientations of symmetry axes of particles $i$ and $j$, the
orientation of the vector joining them is denoted by the unit
vector ${\hat{\bf r}}$. The dependence of the contact distance on the 
orientations of the particles and the interparticle vector is

\begin{widetext}
\begin{equation} 
\sigma({\hat{\bf e_i}}, {\hat{\bf e_j}}, {\hat{\bf r}}) = 
\sigma_0 \left[ 1 - \chi\left(\frac{({\hat{\bf e_i}}.{\hat{\bf r}})^2
+ ({\hat{\bf e_j}}.{\hat{\bf r}})^2 - 2\chi({\hat{\bf e_i}}.
{\hat{\bf r}})({\hat{\bf e_j}}.{\hat{\bf r}})
({\hat{\bf e_i}}.{\hat{\bf e_j}})}
{1 - \chi^2({\hat{\bf e_i}}.{\hat{\bf e_j}})^2}\right)\right]^{-\frac{1}{2}}
\end{equation} 
\end{widetext}
where $\sigma_0$ is the contact distance for the cross configuration
$({\hat{\bf e_i}}.{\hat{\bf e_j}}= {\hat{\bf e_i}}.{\hat{\bf r}}
={\hat{\bf e_j}}.{\hat{\bf r}} = 0) $.
The parameter $\chi$ is a function of the ratio 
$x_0(\equiv \frac{\sigma_e}{\sigma_s})$, 
which is defined in terms of the contact distances when the 
particles are end-to-end (e) and side-by-side (s),

\begin{equation}
\chi = \frac{x^2_0 - 1}{x^2_0 + 1}
\end{equation}

 This vanishes for a sphere and tends to the limiting value of unity 
for an infinitely long rod. The orientational dependence of the 
potential well depth is given by a product of two functions,

\begin{equation}
\epsilon ({\hat{\bf e_i}}, {\hat{\bf e_j}}, {\hat{\bf r}}) =
\epsilon_0 \epsilon^{\nu}({\hat{\bf e_i}}, {\hat{\bf e_j}})
\epsilon^{\prime\mu}({\hat{\bf e_i}}, {\hat{\bf e_j}}, {\hat{\bf r}})
\end{equation}
where the scaling parameter $\epsilon_0$ is the well depth for the cross 
configuration. The first of these functions

\begin{equation} 
\epsilon ({\hat {\bf e_i}}, {\hat {\bf e_j}}) = [1 - \chi^2
({\hat {\bf e_i}}.{\hat {\bf e_j}})^2]^{-\frac{1}{2}}
\end{equation} 
clearly favours the parallel alignment of the particles and so 
aids liquid crystal formation. The second function has a form 
analogous to $\sigma({\hat{\bf e_i}}, {\hat{\bf e_j}}, {\hat{\bf r}})$,
{\it i.e.}

\begin{widetext}
\begin{equation} 
\epsilon'({\hat {\bf e_i}}, {\hat {\bf e_j}}, {\hat{\bf r}}) = 
\left[ 1 - \chi'\left(\frac{({\hat{\bf e_i}}.{\hat{\bf r}})^2
+ ({\hat{\bf e_j}}.{\hat{\bf r}})^2 - 2\chi'({\hat{\bf e_i}}.
{\hat{\bf r}})({\hat{\bf e_j}}.{\hat{\bf r}})
({\hat{\bf e_i}}.{\hat{\bf e_j}})}
{1 - \chi^{\prime 2}({\hat{\bf e_i}}.{\hat{\bf e_j}})^2}\right)
\right]
\end{equation} 
\end{widetext}
where the parameter $\chi'$ is determined by the ratio of
the well depths, $k'(\equiv \frac{\epsilon_s}{\epsilon_e})$,
via 
\begin{equation}  
\chi' = \frac{k^{\prime 1/\mu}-1}{k^{\prime 1/\mu}+1}
\end{equation}

 The potential contains four parameters $(x_0, k', \mu, \nu)$
which determine the anisotropy in the repulsive and attractive forces,
in addition to two parameters $(\sigma_0, \epsilon_0)$ which scale
the distance and energy,respectively. The ratio of the end-to-end 
and side-by-side contact distance, $x_0$, is related to the 
anisotropy of the repulsive forces and it also determines the difference
in the depth of the attractive well between the side-by-side and
the cross configurations.The parameter $k'$ is the ratio of the 
well depth for the side-by-side and end-to-end configurations.
While $x_0$ determines the ability of the system to form an
orientationally ordered phase, $k'$ determines the tendency of the
system to form a smectic phase \cite{7}.The other two parameters 
$\mu$ and $\nu$   
influence nematic and smectic forming character of the
anisotropic attractive forces in a more subtle way.

 In almost all of the simulation and theoretical studies to date 
n has been taken equal to 12. The value of n defines the nature of
the repulsion; the higher the value of $n$ the harder is the nature 
of the repulsion. In Fig.1 we plot $u^*(r, \Omega_1, \Omega_2) 
( = u(r, \Omega_1, \Omega_2)/\epsilon_0)$ 
as a function of separation for some fixed orientations with n=10 
and 18. It shows that as n increases the importance of attractive 
interaction increases for all orientations.
In the present paper we investigate the effect
of variation of n {\it i.e.} variation of the range of repulsion on
the properties of molecular liquids and on its freezing transition.

The paper is organized as follows:
In Sec.II, we describe the solution of the Ornstein-
Zernike equation using the Percus Yevick closure relation for pair
correlation functions.
Section III discusses the essential details of density functional
formalism applied to study the freezing of molecular fluids into 
ordered phases. The results are given and discussed in section IV. 

\section{Pair Correlation Functions: Solution of the Percus-Yevick equation}

The single particle density distribution $\rho(\bn)$ defined as 
\be\
\rho(\bn) = \rho(\br, \bom) = \langle \sum_{i=1}
\delta(\br-\br_i)\delta(\bom-\bom_i)\rangle 
\ee\
where $\br_i$ and $\bom_i$ give the position and the orientation of 
$i^{th}$ molecule, the angular bracket represents the ensemble average
and the $\delta$ the Dirac delta function, is constant independent of
position and orientation for an isotropic fluid. It therefore contains
no information about the structure of the system. The structural information
of an isotropic fluid is contained in the two-particle density
distribution $\rho(\bn,\bp)$ which gives the probability of finding 
simultaneously a molecule in a volume element $d\br_1d\bom_1$ centered
at $(\br_1, \bom_1)$ and a second molecule in a volume element 
$d\br_2 d\bom_2$ centered at $(\br_2, \bom_2)$. $\rho(\bn,\bp)$ is defined
as 

\begin{widetext}
\be\
\rho(\bn,\bp) \equiv \rho(\br_1, \bom_1; \br_2, \bom_2) =  \langle 
\sum_{i\neq j}
\delta(\br_1-\br_i)\delta(\bom_1-\bom_i)\delta(\br_2-\br_j)
\delta(\bom_2-\bom_j)\rangle  
\ee\
\end{widetext}
The pair correlation function $g(\bn,\bp)$ is related to $\rho(\bn,\bp)$ 
by the relation 

\be\
g(\bn,\bp) = \frac{\rho(\bn,\bp)}{\rho(\bn)\rho(\bp)}
\ee\

Since for the isotropic fluid $\rho(\bn) = \rho(\bp) = 
\rho_f = \frac{<N>}{V}$ where $ <N>$ is the average number of molecules in
volume $V$,

\be\
\rho_f^2 g(\br, \bom_1, \bom_2) = \rho(\br, \bom_1, \bom_2) 
\ee\
where $\br = ( \br_2 - \br_1)$.
In the isotropic phase $\rho(\bn,\bp)$ depends only on the distance 
$|\br_2-\br_1|= r$, the orientation of molecules with respect to each 
other and on the direction of vector $\br ({\hat \br} = \br/r$
is a unit vector along $\br$). The pair distribution function 
$g(\bn,\bp)$ of the isotropic fluid is of particular interest as it is the
lowest order microscopic quantity which contains informations about the
translational and the orientational structures of the system and also
has direct contact with intermolecular (as well as with intramolecular)
interactions. For an ordered phase, on the other hand, most of the 
structural informations are contained in $\rho({\bf x})$ (see Sec. III).

The values of the pair correlation functions as a function of 
intermolecular separation and orientations at a given \tm\ and pressure
are found either by computer simulations or by solving the 
Ornstein-Zernike equation

\begin{eqnarray}
h(\bn, \bp) - c(\bn, \bp)= && \gamma(\bn, \bp)  \\ \nonumber
= &&\rho_f\int c(\bn, \bs)[\gamma(\bp, \bs) + c(\bp, \bs)] d\bs 
\end{eqnarray}
where $d\bs = d\br_3 d\bom_3$ and $h(\bn, \bp) = g(\bn, \bp) - 1$ and
$c(\bn, \bp)$ are, respectively, the total and direct pair correlation
functions, using a suitable closure relation. Most commonly used closer
relations are the \PY\  (PY) and the hyper netted chain (HNC) relations.
Approximations are introduced through these closure relations. The PY
and HNC integral equation theories are given by the OZ equation coupled 
with the closure relation \cite{8}

\be\
C^{PY}(\bn, \bp) = f(\bn, \bp)[1+\gamma(\bn, \bp)] 
\ee\
and
\be\
C^{HNC}(\bn, \bp) = h(\bn, \bp) - \ln[1+ h(\bn, \bp)]-\beta u(\bn, \bp) 
\ee\
respectively. Here $f(\bn, \bp) = \exp[-\beta u(\bn, \bp)] - 1$ and
$ \beta = (k_BT)^{-1}$.

Both the PY and HNC integral theories have been used to find the 
pair-correlations functions of model fluids of non-spherical 
molecules \cite{9,10}. It is found that while the PY theory underestimates the
correlations, particularly the angular correlation while the HNC theory
overestimates them. In case of hard-core fluids we proposed a 'mixed'
integral equation which interpolates between the HNC and PY theories
and is thermodynamically consistent \cite{11}. Such an approach is needed for 
the soft-core potential the one considered in this paper also.
We, however, defer this approach for the future and confine ourselves here 
to solve the PY equation to get the pair correlation functions for the
GB(n-6) potential.

The angle dependent function $A({\bf r_{12}, \Omega_1, \Omega_2})$ 
(where A may be pair correlation function or pair potential)
is expanded in a basis set of rotational invariants \cite{8}
in space fixed (SF) frame according to the equation

\begin{widetext}
\begin{equation}
A({\bf r_{12}, \Omega_1, \Omega_2}) = \sum_{l_1 l_2 l} \sum_{m_1 m_2 m}
A_{l_1 l_2 l}(r_{12})C_g(l_1 l_2 l;m_1 m_2 m) Y_{l_1 m_1}({\bf\Omega_1})
Y_{l_2 m_2}({\bf\Omega_2})Y^*_{l m}({\bf \Omega})
\end{equation}
\end{widetext}
where $C_g(l_1 l_2 l;m_1 m_2 m) $ are the Clebsch-Gordon coefficients.

 For fully axially symmetric particles it is also possible to expand
the function in products of spherical harmonics in body fixed (BF)
frame according to the equation.

\begin{equation} 
A({\bf r_{12}, \Omega_1, \Omega_2}) = \sum_{l_1 l_2 m}A_{l_1l_2m}(r_{12})  
Y_{l_1 m}({\bf \Omega_1}) Y_{l_2 {\underline m}}({\bf \Omega_2})
\end{equation} 
where ${\underline m} \equiv -m$. Numerically it is easier to 
calculate the BF harmonic coefficients than the SF harmonic coefficients.
The two harmonic coefficients are related through a linear 
transformation,

\begin{displaymath}
A_{l_1l_2m}(r_{12}) = \sum_l \left(\frac{2l+1}{4\pi}\right)^{1/2}
                      A_{l_1 l_2 l}(r_{12})C_g(l_1 l_2 l; m 
{\underline m} 0)
\end{displaymath}
or
\begin{equation}
A_{l_1l_2l}(r_{12}) = \sum_m \left(\frac{4\pi}{2l+1}\right)^{1/2}
                      A_{l_1 l_2 m}(r_{12})C_g(l_1 l_2 l; m 
{\underline m} 0) 
\end{equation}

 In any numerical calculation we can handle only a finite
number of the spherical harmonic coefficients for each 
orientation-dependent function. The accuracy of the results depends
on this number. As the anisotropy in the shape of molecules (or
in interactions) and the value of fluid density $\rho_f$ increases
more harmonics are needed to get proper convergence. We have found 
that the series get converged if we truncate the series
at the value of $l$ indices equal to 6 for molecules with $x_0 \leq 3$ 
\cite{9}. Though it is desirable to include higher order
harmonics {\it i.e.} for $l > 6$ but it will increase computational
time many fold. Our interest is to use the data of the harmonics
of pair correlation functions for freezing transitions where only low
order harmonics are generally involved (see Sec. III below).
The only effect the higher-order
harmonics appear to have on these low-order harmonics is
to modify the finer structure of the harmonics at small values
of $r$ whose contributions to the structural parameters 
(to be define below) are negligible.

 Using the numerical procedure outlined elsewhere \cite{9},
we have solved the PY equation for the GB(n-6) fluid having $n$ values
8, 10, 12, 14, 16, 18, 20, 22, 24, 26, 28 and 30 for $x_0 = 3.0$
and well depth ratio $k' = 5$ at reduced temperatures, $T^* = 
\frac{kT}{\epsilon_0} = 0.65, 0.80, 0.95 $ and $1.25$  for a wide 
range of densities. The other two parameters $\mu$ and $\nu$ are taken
to be $2$ and $1$, respectively.
The solutions could be found only upto certain density $\rho'$
the value of which depend upon the \tm\ and the value of $n$. The value
of $\rho'$ is often close to the isotropic-nematic transition. Because
of this one faces problems in locating other less symmetric phases
of the system using the theory to be discussed
in Sec. III.

 In Fig.2 we compare the values of $g(r) = 1+\frac{h_{000}(r)}{4\pi}$
in BF frame at $T^* = 0.80$ and density $\eta(\equiv \frac{\pi}{6}
\rho_f\sigma^3_0x_0) = 0.25$ for four sets of (n-6) combinations. 
It is seen from this figure that the first peak becomes sharper
and attains its maximum value at smaller
value of $r^*(=\frac{r}{\sigma_0})$ as the hardness of the core increases.
The cause of this becomes clear if we look at Figs.3 and 4 which depict
$v(r) = -T^*\ln[\langle e^{-\beta u(r, \Omega_1, \Omega_2)}
\rangle_{\Omega_1, \Omega_2}$ ] as a function of interparticle separation 
at $T^*=0.8 \;\; {\rm and} 1.25$, respectively.  
$v(r)$ may be regarded as an averaged pair potential and,
therefore, helps us in understanding the features of $g(r)$.
$v(r)$ seems to have two minimum; one at $r^* \approx 1.25$
and other at $r^* \approx 2.25$. The first minimum becomes deeper
at higher $n$ and at lower \tm\ and almost vanishes at lower $n$
and higher temperatures. The second minimum dependence on $n$ 
(as well as on \tm\ ) is weak. One may also note the shift
to lower values of $r^*$ of first minimum as $n$ is increased. 

Since the PY theory is known to be reasonably accurate for systems
interacting via pair potential which has hard repulsive core 
$(n \to \infty)$ and weak attraction \cite{11}, the values of the pair
correlation functions reported here are expected to be more accurate
for higher values of $n$ and lower values of $T^*$ compared to values
corresponding to lower $n$ and higher $T^*$.

 In Figs.5-6 we compare the two other projections of PCF in BF-
frame at the same state conditions and observe similar behavior. 
In Fig.7 we compare the value of $g(r)$ at $\eta = 0.5$
for GB(10-6) model at four different temperatures. Here we see that
the first peak gets sharper as the temperature decreases. Such behavior
is also seen (see Fig. 2) when $n$ is increased at the same temperature. 
This is due to increasing tendency of the molecules to form 
parallel configurations.

As has already been mentioned, the PY theory underestimates the 
molecular correlations. This can be seen from the pressure calculated 
using the values of the
direct pair correlation function through the compressibility relation
which is found to be lower than the simulated value as shown in Fig.8.

For a system consisting of axially symmetric non-dipolar molecules
the static Kerr constant $K$ is given by \cite{12,13}

\bm\
K = \beta \kappa \left[1-\frac{{\hat C}^0_{22}}{5}\right]^{-1}
\em
where ${\hat C}^0_{22}$ is structural parameter defined as in Eq.(3.22)
and $\kappa$ is a constant dependent only upon single particle 
properties. The divergence of  $K$ may signal the absolute stability
limit of the isotropic phase relative to orientationally ordered phase 
\cite{12}.
Thus the isotropic phase becomes orientationally unstable when the
inverse Kerr constant $K^{-1} \to 0$. It is, however, important to 
emphasize that the condition $K^{-1} \to 0$ does not determine 
the thermodynamic phase transition, but rather a point on the spinodal 
line. This means that the density at which $K^{-1} = 0$ establishes 
a stability limit in the sense that at higher densities the isotropic 
phase cannot exist even as a metastable state.

The reduced Kerr constants $\beta A K^{-1}$ as a function of $\eta$
for the various $n$ values are plotted in Figs. 9 \& 10 at $T^* = 0.8 $
and $1.25$.

\section{Theory for freezing}

The structural informations of fluids at the \pcf\ level obtained
above can be used to obtain information about their freezing. At the
freezing point the spatial and orientational configurations of
molecules undergo a modification. Often abrupt change in the symmetries
of the system takes place on the freezing. In contrast to the isotropic
fluid, the molecular configurations of most ordered phases are 
adequately described by the single particle density (singlet)
distribution $\rho({\bf x}).\rho({\bf x})$ provides us with a convenient
quantity to specify an arbitrary state of a system. One may consider
a variational thermodynamic potential as a functional of  $\rho({\bf x})$.
The equilibrium state of the system at given $T$ and $P$ is described by
the density $\rho(T,P,{\bf x})$ corresponding to the minimum
of the thermodynamic potential with respect to $\rho({\bf x})$. This
forms the basis of the density functional theory.

In this article we investigate the freezing of the GB(n-6) fluid
into the nematic and the smectic A (Sm A) phases using density functional 
theory (DFT). 
In the nematic phase the full translational symmetry of the isotropic
fluid phase (denoted as $R^3$) is maintained but the rotational symmetry
$O(3)$ or $SO(3)$ (depending upon the presence or absence of the centre
of symmetry) is broken. In the simplest form of the axially symmetric 
molecules the group $O(3)$ (or  $SO(3)$) is replaced by one of the
uniaxial symmetry $D_{\infty h}$ (or $D_{\infty}$). The phase possessing
the $R^3 \wedge D_{\infty h}$ (denoting the semi direct product of the
translational group $R^3$ and the rotational group $D_{\infty h}$)
symmetry is known as uniaxial nematic phase \cite{1,14}.

The smectic liquid crystals, in general, have a stratified structure with
th long axes of molecules parallel to each other in layers. This situation
corresponds to partial breakdown of translational invariance in addition
to breaking of the orientational invariance. Since a variety of 
molecular arrangement are possible within each layer, a number of smectic
phases are possible \cite{1}. The simplest among them is the Sm A phase.
In it the centre of mass of molecules in a layer are distributed 
as in a two-dimensional fluid but the molecular axes are on the average
along a direction normal to the smectic layer ({\it i.e.} the director
${\hat n}$ is normal to the smectic layer). The symmetry of the Sm A phase is
$D_{\infty h} \wedge (R^2 \times Z)$ where $R^2$ corresponds to a 
two-dimensional liquid structure and $Z$ for a one-dimensional periodic
structure.

The order parameters which characterize the ordered structures can be
found from the singlet distribution $\rho(x)$. For this we express it in the
Fourier series and the Wigner rotation matrices. Thus

\be\
\rho({\bf x}) = \rho(\br, \bom) = \rho_0 \sum_q \sum_{lmn} Q_{lmn}
(G_q) \exp(i{\bf G}_q.\br) D^l_{mn}(\bom)
\ee\ 
where the expansion coefficients
\be\
 Q_{lmn}(G_q) = \frac{2l+1}{N}\int d\br \int d\bom \rho(\br, \bom)
\exp(-i{\bf G}.\br) D^{l*}_{mn}(\bom)  
\ee\
are the order parameters, $G_q$ the reciprocal lattice vectors, $\rho_0$
the mean number density and $D^l_{mn}(\bom)$ the generalized spherical
harmonics or Wigner rotation matrices \cite{15}.

Since we are interested in uniaxial systems of cylindrically symmetric 
molecules, $m=n=0$ in Eqs.(3.1) \& (3.2). This leads to 
\be\
\rho(\br, \bom) = \rho_0 \sum_l \sum_q Q_{lq} \exp(i{\bf G}_q.\br) 
                  P_l(\cos \theta)
\ee\   
and
\be\
Q_{lq} = \frac{2l+1}{N}\int d\br \int d\bom \rho(\br, \bom)
\exp(-i{\bf G}.\br) P_l(\cos \theta) 
\ee\     
where $ P_l(\cos \theta)$ is the Legendre polynomial of degree $l$ and 
$\theta$ is the angle between the cylindrical axis of a molecule and the
director.

Since in the nematic phase the centres of mass of molecules are distributed
as randomly as in the isotropic fluid but the molecular axes are aligned
along a particular direction defined by the director ${\hat{\bf n}}$
(a unit vector) we have $G_q=0$ and
\be\
Q_{l0} = \langle(2l+1) P_l(\cos(\theta))\rangle = (2l+1){\bar P_l}
\ee\ 
where angular bracket indicates the ensemble average. It is often
enough to use two orientational order parameters ${\bar P_2}$ and
${\bar P_4}$ to locate the isotropic-nematic transition as in almost
all known cases the transition is weak first-order transition \cite{1}.

To characterize the Sm A phase we need three different class of order 
parameters; (i) orientational, (ii) positional, (iii) mixed. These
parameters are found from Eq.(3.2). For the orientational order we take
${\bar P_2}$ and ${\bar P_4}$ as in case of the nematic phase. For
the positional order along the z-axis we choose one order parameter
corresponding $G_z = \frac{2\pi}{d}$, $d$ being the layer spacing. Thus

\be\
\mu = Q_{00}(G_z) = \langle \cos( \frac{2\pi z}{d})\rangle
\ee\
The coupling between the positional and orientational ordering is described
by the (mixed) order parameter $\tau$ defined as
\be\
\tau = \frac{1}{5} Q_{20}(G_z) = \langle \cos (\frac{2\pi z}{d})
P_2(\cos \theta)\rangle  
\ee\

We therefore choose four order parameters to describe the ordering
in a Sm A phase and two for the nematic ordering. Another way of writing
the trial singlet distribution corresponding to the ordered phases of
our interest is

\begin{widetext}
\be\
\rho(\br,\bom) = A_0 \rho_0 \exp[-\alpha(z-d)^2 - \alpha^1(z-d)^2 
P_2(\cos \theta) + \lambda_2 P_2(\cos \theta) + \lambda_4 P_4(\cos \theta)]
\ee\
\end{widetext}
where $A_0$ is a normalization constant, $\alpha$ and $\alpha^1$ are
associated with the formation of layer in the Sm A phase and $\lambda_2$ 
and $\lambda_4$ with orientational ordering. When  $ \alpha$ and $ \alpha^1$ 
are zero but  $\lambda_2$ and $\lambda_4$ are non zero the phase is
nematic. In case of the isotropic fluid all the four parameters  $ \alpha$,
$ \alpha^1$, $\lambda_2$ and $\lambda_4$ are zero.
If all the four parameters are non zero the phase is Sm A. The
four order parameters defined above can be found taking the expression 
of $\rho(\br, \bom)$ given by Eq.(3.8). Thus
\ba\
\mu & = & \frac{A_0}{d}\int_0^d dz \cos (\frac{2\pi z}{d})
          \int_0^1 dx \exp(S) \\
\tau & = &  \frac{A_0}{d}\int_0^d dz \cos (\frac{2\pi z}{d})
          \int_0^1 dx \exp(S) P_2(x)\\ 
{\bar P}_2 & = &  \frac{A_0}{d}\int_0^d dz \int_0^1 dx \exp(S) P_2(x)\\
{\bar P}_4 & = &  \frac{A_0}{d}\int_0^d dz \int_0^1 dx \exp(S) P_4(x)
\ea\
where $S = -\alpha(z-d)^2 - \alpha^1(z-d)^2 P_2(x) + \lambda_2 P_2(x) 
          + \lambda_4 P_4(x)$  

\subsection{Density Functional Approach}

In the usual \dft\ aapproach one uses the grand thermodynamic potential
to locate the transition. The grand thermodynamic potential is defined as

\begin{equation}
-W = \beta A - \beta \mu \int d{\bf x} \rho({\bf x})
\end{equation}
where A is the Helmholtz free energy, $\mu$ the chemical potential and 
$\rho({\bf x})$ is a singlet distribution function. Eq.(3.1) can be written as

\begin{equation}
\Delta W = W - W_f = \Delta W_1 + \Delta W_2
\end{equation}
where $W_f$ is the grand thermodynamic potential of the isotropic 
fluid, and \cite{14}.

\begin{widetext}
\begin{eqnarray}
\frac{\Delta W_1}{N} & = & \frac{1}{\rho_f V} \int {d\bf r}
{d\bf \Omega}\left\{{\rho({\bf r}, {\bf \Omega}) 
\ln \left[\frac{\rho({\bf r}, {\bf \Omega})}{\rho_f}\right] - \Delta 
\rho({\bf r}, {\bf \Omega})}\right\} \\
{\rm and} \nonumber \\
\frac{\Delta W_2}{N} & = & -\frac{1}{2\rho_f} \int {d\bf r_{12}}
{d\bf \Omega_1}{d\bf \Omega_2}\Delta\rho({\bf r_1}, {\bf \Omega_1}) 
c({\bf r_{12}}, {\bf \Omega_1}, {\bf \Omega_2})\Delta\rho({\bf r_2}, 
{\bf \Omega_2}) 
\end{eqnarray}
\end{widetext}  
Here $\Delta\rho({\bf x}) = \rho({\bf x}) - \rho_f$, where 
$\rho_f$ is the density of the coexisting liquid.
The ordered phase density is found by minimizing $\Delta W$ with
respect to arbitrary variations in the ordered phase density subject
to the constraint which corresponds to some specific features of the ordered
phase. Thus,

\begin{equation} 
\ln\frac{\rho({\bf r_1},{\bf \Omega_1})}{\rho_f} = \lambda_L + 
\int d{\bf r_2} d{\bf \Omega_2} c({\bf r_{12}}, {\bf \Omega_1}, 
{\bf \Omega_2};\rho_f)\Delta\rho({\bf r_2}, {\bf \Omega_2})
\end{equation}
where $\lambda_L$ is Lagrange multiplier which appears in the
equation because of constraint imposed on the minimization.

One attempts to find solution of $\rho({\bf x})$ of Eq.(3.17) which have
symmetry of the ordered phase. These solutions, inserted in Eq.(3.14)
give the grand thermodynamic potential difference between the ordered
and liquid phases. The phase with the lowest grand potential is
taken as the stable phase. Phase coexistence occurs at the value
of $\rho_f$ which makes $-\frac{\Delta W}{N}=0$ for the ordered
and liquid phases. Substituting Eq.(3.1), into Eq.(3.17) and Eq.(3.14)
and integrating results in, respectively 

\begin{widetext}
\begin{equation}
\delta_{l'0} \delta_{q'0} + \frac{Q_{l'q'}}{2l'+1}  =  
\frac{1}{V}\int d{\bf r_1}d{\bf \Omega_1}e^{-i{{\bf G}_{q'}}.{\bf r_1}}
P_{l'}(\cos \theta_1) \exp[\lambda_L  +  \sum_l\sum_q \frac{Q_{lq}}{2l+1} 
e^{-i{{\bf G_{q}}}.{\bf r_1}}{\hat C}^q_{l,0}(\theta_1)] 
\end{equation} 
\end{widetext}
and
\begin{eqnarray}
-\frac{\Delta W}{N} & = & -\Delta \rho^* + 
\Delta \rho^*{\hat C}^0_{0,0} \nonumber \\
& & + \frac{1}{2}\sum_{LL'}\sum_q \frac{Q_{Lq} Q_{L'q}}{(2L+1)(2L'+1)}
{\hat C}^q_{L,L'} \nonumber \\
\end{eqnarray} 
where 

\begin{eqnarray}
Q_{0,0} &=&\Delta \rho^* \\
{\hat C}^q_{l,0}(\theta_1)&=&(2l+1) \rho_f \int 
d{\bf r_{12}}d{\bf \Omega_2}c(\br_{12}, \bom_1, \bom_2) \\ \nonumber
&&e^{i{\bf G}_q.{\bf r_{12}}}
P_{l}(\cos \theta_2)  \\
{\hat C}^q_{l,l'}&=& (2l+1)(2l'+1)\rho_f \int
d{\bf r_{12}}d{\bf \Omega_1}d{\bf \Omega_2} \\ \nonumber
&&e^{i{\bf G}_q {\bf r_{12}}}c(\br_{12}, \bom_1, \bom_2)
P_{l}(\cos \theta_1) P_{l'}(\cos \theta_2)                      
\end{eqnarray}
are the structural parameters related to the Fourier transformed
direct correlation function of the fluid phase. Eq.(3.18) is the 
expression for the order parameters. This version of the \dft\ is known
as the second order density functional (SODFT) because it considers only the
\pcf\ and neglects the higher order correlations which might be present
in the system at the transition point.

\subsection{Modified Weighted-Density Approximation (MWDA)}

In another version of the density functional approach in which higher order
correlations are included and known as Modified Weighted Density 
Approximation \cite{16}, one uses the Helmholtz free energy to locate
the transition. For the Helmholtz free energy we write 

\be\
A[\rho(\br, \bom)] = A_{id}[\rho(\br, \bom)] + A_{ex}[\rho(\br, \bom)]
\ee\
where both terms in Eq.(3.23) are unique functionals of the one-particle
density $\rho(\br, \bom)$. The first term in the right hand side
of Eq.(3.23) is a non uniform ideal gas contribution of the form

\be\
A_{id}[\rho(\br, \bom)] = \beta^{-1}\int_V d\br d\bom \rho(\br, \bom)
                          \{\ln[\rho(\br, \bom) \lambda^3] - 1\}
\ee\
where $\lambda$ is the thermal de Broglie wavelength. The second term in
the right hand side of Eq.(3.23) is the excess Helmholtz free energy
of the non uniform system.

In the modified weighted-density approximation the excess free
energy of a uniform system, but evaluated of a weighted density 
${\hat\rho}$ \cite{16}

\be\
A^{MWDA}_{ex}[\rho] = N \phi_0({\hat \rho})
\ee\
where $N$ is the number of particles in the system $\phi_0(\rho)$
is the excess free energy per particle of a uniform system at 
density $\rho$. The weighted density ${\hat \rho}$ is constructed
from the actual inhomogeneous one-particle density $\rho({\bf x})$ and
is defined by

\be\
{\hat \rho} = \frac{1}{N} \int_V d{\bf x}\rho({\bf x})\int_V d{\bf x}'
                \rho({\bf x}'){\tilde\omega}({\bf x}-{\bf x}';{\hat \rho})
\ee\
introducing thereby the weighted function 
${\tilde\omega}({\bf x}-{\bf x}';{\hat \rho})$. It is an essential 
ingredient of the MWDA that the weighted function ${\tilde\omega}$
which is used to determine the weighted density, depends itself on the
sought function ${\hat \rho}$; thus Eq.(3.26) has to viewed as a
self-consistency condition for the determination of the weighted
density. To ensure that the approximation in the determination of 
${\hat \rho}$ becomes exact in the uniform limit, the weighted function
has to be normalized, {\it i.e.},

\be\
\int d{\bf x} {\tilde\omega}({\bf x}-{\bf x}';{\hat \rho}) = 1
\ee\
for any ${\hat \rho}$. The function ${\tilde\omega}$ can be then
uniquely specified by requiring that the approximate functional
$A^{MWDA}_{ex}[\rho]$ is exact upto second order in the functional
expansion, namely

\be\
C({\bf x}-{\bf x}';\rho_0) = -\beta \lim_{\rho\to \rho_0}
\left[\frac{\delta^2 A^{MWDA}_{ex}[\rho]}{\delta \rho({\bf x})
\delta \rho({\bf x}')}\right]   
\ee\

The conditions [Eqs.(3.25-3.28)] result in a particularly simple
expression for ${\tilde\omega}$, namely

\be\
{\tilde\omega}({\bf x}-{\bf x}';{\hat \rho}) = 
-\frac{1}{2\phi'_0({\hat \rho})}\left[ \beta^{-1} 
C({\bf x}-{\bf x}';{\hat \rho}) + \frac{1}{V}
{\hat \rho}\phi_0^{''}({\hat \rho}) \right]
\ee\
where $V$ is the volume of the sample, $\phi_0({\hat \rho})$ is the
excess free energy per particle of an isotropic fluid of density
${\hat \rho}$ and primes on $\phi_0({\hat \rho})$ indicate
derivatives with respect to density. Using expansion (Eq. 3.1) and
(Eq.2.8), respectively for $\rho({\bf x})$ and 
$C({\bf x}-{\bf x}';{\hat \rho})$ we find for the ordered phase

\begin{eqnarray}
{\hat \rho} &=& \rho_0 \sum_{L_1} \sum_{L_2}  \sum_q
Q_{L_1q}Q_{L_2q} \frac{{\hat c}_{L_1 L_2}^q}{(2L_1+1)(2L_2+1)} \nonumber \\
&&\left[-\frac{1}{2{\hat \rho}\beta \phi_0^{'}({\hat \rho})} \right]
-\rho_0{\hat \rho}\frac{\phi_0^{''}({\hat \rho})}{2\phi_0^{'}({\hat \rho})}
\end{eqnarray}

Having computed ${\hat \rho}$, the next step in freezing analysis
is to substitute ${\hat \rho}$ into Eq.(3.25) to compute
$A^{MWDA}_{ex}$. In terms of structural parameter, the excess free
energy per particle of a uniform system at a density $\rho$ is given
as 

\be\
\beta \phi_0(\rho) = -\int_0^{\rho} d\rho{''} \frac{1}{\rho{''2}}
\int_0^{\rho{''}} {\hat C}^0_{00}[\rho'] d\rho'
\ee\ 

The ideal gas part is calculated using the ansatz for $\rho(\br, \bom)$
given by Eq.(3.8). Thus

\begin{eqnarray}
\beta A_{id}[\rho_0] &=& \rho_0 \int d\br \sum_L \sum_q
\frac{Q_{Lq}}{\sqrt {2L+1}} e^{i{\bf G}_q.\br}  
[\{\ln(A_0\rho_0\lambda^3)-1\} \delta_{L0}
-\alpha(z-x_0)^2\frac{\delta_{ 2L}}{\sqrt 5}  \nonumber \\
&& +  \frac{\lambda_2\delta_{2L}}{\sqrt 5} +  
\frac{\lambda_4\delta_{4L}} 3
- \alpha^1(z-x_0)^2 \delta_{L0}] 
\end{eqnarray}

To determine the transition parameters, we first compute the effective 
density ${\hat \rho}$ from Eq.(3.30) and minimizing the free energy from
Eqs.(3.23, 3.31 and 3.32) with respect to $\rho_0, \alpha, \lambda_2, 
\lambda_4 \;\;
{\rm and}\;\; \alpha^1$. In order to determine the transition density of the
coexisting isotropic $(\rho_f)$ and anisotropic $(\rho_0)$ phases it is
necessary to equate the pressure and chemical potentials 
(Maxwell construction) of the two phases.

\section{Results and Discussion}

We have used both versions of the density functional methods described 
above to locate the freezing transitions and calculate the values of the 
freezing parameters. The structural parameters defined by Eq.(3.22) which 
appear in the \dft\ as the input data are obtained from the harmonics of 
the direct \pcf\ evaluated using the PY integral equation theory 
(given in Sec. II). Using these values of the structural parameters and 
the four order parameters ${\bar P_2},{\bar P_4}, \mu \;\; {\rm and} 
\tau$ we have solved Eqs.(3.18-3.19) of the SODFT and Eqs.(3.23-3.32) 
of the 
MWDA for the GB(n-6) fluid with $8 \leq n \leq 30$ for temperatures
lying between $0.65$ to $1.25$. All our results correspond to 
$\mu = 2, \nu = 1, x_0 = 3 \;\; {\rm and}\;\; k' = 5$.  

Our results show that for none of the cases studied here Sm A phase gets 
stabilized. In the low temperature region for a given $n$ it, however, 
appeared as a metastable state having free energy lower than that of the
isotropic phase but higher than the nematic (see Table I). Since we have 
not included Sm B and crystalline phases in our investigation for the reason
already given, we found only the isotropic-nematic transition.

For each $n$ we found a lower cut-off of the \tm\ for the existence of 
the nematic phase. The nematic phase was not found to exist below this
temperature. The lower cut-off \tm\ for the nematic phase is found to increase
with $n$. For example, where for $n = 8 \;\;{\rm and}\;\;10$ we found the 
nematic phase
to exist at $T^* = 0.65$ but not for $n \geq 12$. The computer simulation
results of Miguel {\it et al} \cite{5} show that for $n = 12$ the 
cut-off \tm\ is slightly above $T^* = 0.8$. Our results, however, show that
the nematic phase exists at $T^* = 0.8$. This may be due to error in the
structural parameters values found from the PY theory. 

Both versions of the \dft\ give similar
results for the transition density $\rho^*_f$ but give the values of the
order parameters including the change in density at the transition
which are different from each other. More surprising is the way the values 
of the order parameters ${\bar P_2}$ and ${\bar P_4}$ vary with 
\tm\ and with $n$ (see Tables II-V) found from the two version of 
the theory. While the SODFT predicts that  ${\bar P_2}$ and ${\bar P_4}$ 
decrease as the transition \tm\ is increased, the MWDA predicts them 
to increase. The computer simulation results \cite{5,6} do not give any 
clear indication as how these parameters vary with transition temperature. 
Similar difference in the variation of the values of the order parameters 
with $n$ is also found.

In Table II-V we give the values of the transition parameters found from
the two theories. We also give the results found from the computer 
simulations at $T^* = 0.95 \;\; {\rm and}\;\; 1.25$ for $n=12$. There is
very good agreement between these results at $T^* = 0.95$. The transition
density found from the theories are identical though somewhat higher 
than the value found from the simulation. The value of $\Delta \rho^*$
found from these methods are also in good agreement, though MWDA predicts
the value of  $\Delta \rho^*$ which is lower than the SODFT as well as
simulation value. Pressure and chemical potentials are in good agreement.
But there is difference in the value of  ${\bar P_2}$ and ${\bar P_4}$.
At  $T^* = 1.25$ both theories predict the transition density which is
high compared to the MD value. One of the possible reasons for this is, as
pointed out in Sec. II, the inaccuracy in the values of c-harmonics at
higher temperature. The PY theory is known to underestimate
the angular correlations and this defect of the PY theory becomes more 
pronounced as \tm\ is increased for a given  $n$. This may be the reason
why the theory predicts the transition at higher density than the MD
value. As a consequence of this the transition pressure and the chemical
potential are also substantially higher than the MD values. This comparison
at  $T^* = 0.95 \;\;{\rm and} \;\; 1.25$ show that while the DFT is good to
predict the freezing parameters, the PY values of structural parameters
at higher \tm\ are lower than the actual values.

We hope to combine the PY and HNC theories to generate accurate values
of the harmonics of the \pcf\ and with these values to compute the
full phase diagram. 

\section{acknowledgements}

The work was supported by the Department of Science 
and Technology (India) through project grant. One of us (RCS) thanks
the Director and Management for providing some computational facility
at MIET.

\newpage

\newpage
\begin{widetext}
\begin{table*}
\caption{Values of order parameters and energy of smectic A and nematic 
phases at $T^* = 0.8$ for GB(12-6) potential. While nematic is a stable
phase, smectic A is metastable as its energy is higher than the nematic}
\label{tab1}
\begin{ruledtabular}
\begin{tabular}{cccccccc}
$\rho^*_f$ & Phase & $\mu_z$ & ${\bar P_2}$ & ${\bar P_4}$ & $\tau_{2z}$ & 
$\Delta\rho^*$ & $\Delta W$ \\  \hline
\\
0.293& Sm-A & 0.674 & 0.891 & 0.719 & 0.687 & 0.098 & -0.023 \\
 & Nematic & 0.000 & 0.843 & 0.562 & 0.000 & 0.058 & -0.093 \\
\\
0.306 & Sm-A & 0.647 & 0.888 & 0.741 & 0.677 & 0.088 & -0.157 \\
 & Nematic & 0.000 & 0.903 & 0.654 & 0.000 & 0.063 & -0.243 \\
\\
0.312 & Sm-A & 0.629 & 0.887 & 0.750 & 0.667 & 0.083 & -0.234 \\
 & Nematic & 0.000 & 0.923 & 0.691 & 0.000 & 0.065 & -0.333 
\end{tabular}
\end{ruledtabular}
\end{table*}
\end{widetext}

\begin{widetext}
\begin{table}
\caption{Isotropic-Nematic transition parameters for GB(n-6) fluid
at $T^* = 0.65$. The reduced units are $P^* = P\sigma_0^3/\epsilon_0,
\mu^* = \mu/\epsilon_0, \;\; {\rm and} \;\; \rho^* = \rho\sigma_0^3$ 
}
\label{tab2}
\begin{ruledtabular}
\begin{tabular}{ccccccccc}
Potential Model & Theory & $\rho^*_f$ & $\rho^*_n$ & $\Delta\rho^*$ & 
${\bar P_2}$ & ${\bar P_4}$ & $P^*$ & $\mu^*$ \\   \hline
\\
 (8,6) & DFT & 0.428 & 0.431 & 0.006 & 0.69 & 0.40 & 9.48 & 26.79\\
 & MWDA & 0.412 & 0.416 & 0.009 & 0.56 & 0.27 & 7.77 & 22.71 \\
\\
(10, 6) & DFT & 0.29 & 0.301 & 0.038 & 0.72 & 0.41 & 1.30 & 3.96 \\
 & MWDA & 0.286 & 0.29 & 0.013 & 0.36 & 0.12 & 1.22 & 3.69 \\
\end{tabular}
\end{ruledtabular}
\end{table}
\end{widetext}

\begin{widetext}
\begin{table}
\caption{Same as in Table II but at $T^* = 0.80$ }
\label{tab3}
\begin{ruledtabular}
\begin{tabular}{ccccccccc}
Potential Model & Theory & $\rho^*_f$ & $\rho^*_n$ & $\Delta\rho^*$ & 
${\bar P_2}$ & ${\bar P_4}$ & $P^*$ & $\mu^*$ \\  \hline
\\
(10, 6) & DFT &  0.341 & 0.346 & 0.015 & 0.68 & 0.37 & 3.81 & 12.44 \\
 & MWDA & 0.339 & 0.341 & 0.007 & 0.40 & 0.15 & 3.71 & 12.16 \\
\\
(12, 6) & DFT &  0.282 & 0.295 & 0.046 & 0.74 & 0.43 & 1.38 & 4.11 \\
 & MWDA & 0.277 & 0.281 & 0.015 & 0.36 & 0.12 & 1.27 & 3.69 \\
\\
(14, 6) & DFT &  0.239 & 0.277 & 0.158 & 0.92 & 0.62 & 0.58 & 0.65 \\
 & MWDA & 0.237 & 0.241 & 0.019 & 0.27 & 0.08 & 0.55 & 0.54 \\
\end{tabular}
\end{ruledtabular}
\end{table}
\end{widetext}

\begin{widetext}
\begin{table}
\caption{Same as in Table II but at $T^* = 0.95$ }
\label{tab4}
\begin{ruledtabular}
\begin{tabular}{ccccccccc}
Potential Model & Theory & $\rho^*_f$ & $\rho^*_n$ & $\Delta\rho^*$ & 
${\bar P_2}$ & ${\bar P_4}$ & $P^*$ & $\mu^*$ \\  \hline
\\
(10, 6) & DFT &  0.381 & 0.385 & 0.009 & 0.68 & 0.38 & 8.26 & 25.63 \\
 & MWDA & 0.379 & 0.382 & 0.007 & 0.46 & 0.20 & 8.04 & 25.04 \\
\\
(12, 6) & MD & 0.308  & 0.314 & 0.019 & 0.50 & - & 3.50 & 12.70 \\
 & DFT & 0.322 & 0.328 & 0.02 & 0.67 & 0.37 & 3.40 & 11.28 \\
 & MWDA & 0.322 & 0.325 & 0.008 & 0.37 & 0.13 & 3.40 & 11.28 \\
\\
(14, 6) & DFT & 0.287 & 0.299 & 0.042 & 0.74 & 0.43 & 1.82 & 5.66 \\
 & MWDA & 0.283 & 0.288 & 0.017 & 0.37 & 0.12 & 1.69 & 5.21 \\
\\
(16, 6) & DFT & 0.261 & 0.283 & 0.085 & 0.82 & 0.51 & 1.06 & 2.59 \\
 & MWDA & 0.245 & 0.251 & 0.027 & 0.36 & 0.12 & 0.82 & 1.64 \\
\end{tabular}
\end{ruledtabular}
\end{table}
\end{widetext}

\begin{widetext}
\begin{table}
\caption{Same as in Table II but at $T^* = 1.25$ }
\label{tab5}
\begin{ruledtabular}
\begin{tabular}{ccccccccc}
Potential Model & Theory & $\rho^*_f$ & $\rho^*_n$ & $\Delta\rho^*$ & 
${\bar P_2}$ & ${\bar P_4}$ & $P^*$ & $\mu^*$ \\ \hline
\\
(10, 6)\footnotemark[1] & DFT & 0.454 & 0.456 & 0.005 & 0.72 & 0.44 
& 26.93 & 72.59 \\
 & MWDA & 0.435 & 0.437 & 0.005 & 0.62 & 0.30 & 21.26 & 59.84 \\
\\
(12, 6) & MD[5] & 0.323 & 0.331 & 0.025 & 0.50 & - & 5.70 & 20.90 \\
 & DFT & 0.378 & 0.382 & 0.009 & 0.68 & 0.38 & 10.90 & 34.27 \\
 & MWDA & 0.375 & 0.378 & 0.007 & 0.47 & 0.21 & 10.42 & 32.99 \\
\\
(14, 6) & DFT & 0.344  & 0.349 & 0.014 & 0.68 & 0.38 & 6.57 & 21.88 \\
 & MWDA &  0.343 & 0.346 & 0.009 & 0.44 & 0.20 & 6.52 & 21.71 \\
\\
(18, 6) & DFT & 0.306 & 0.315 & 0.028 & 0.72 & 0.41 & 3.43 & 11.52 \\
 & MWDA & 0.303 & 0.307 & 0.014 & 0.41 & 0.18 & 3.25 & 10.95 \\
\\
(24, 6) & DFT & 0.273 & 0.291 & 0.065 & 0.79 & 0.49 & 1.81 & 5.35 \\
 & MWDA & 0.267 & 0.274 & 0.026 & 0.37 & 0.13 & 1.65 & 4.78 \\
\\
(30, 6) & DFT & 0.249 & 0.283 & 0.137 & 0.90 & 0.61 & 1.11 & 2.38 \\
 & MWDA & 0.242 & 0.248 & 0.027 & 0.34 & 0.11 & 0.99 & 1.90 \\
\end{tabular}
\end{ruledtabular}
\footnotetext[1]{The results have been found by extrapolating
the data of the structural parameters to high densities. The value of the
transition parameters may, therefore, not be as accurate as for the other
cases.}
\end{table}
\end{widetext}

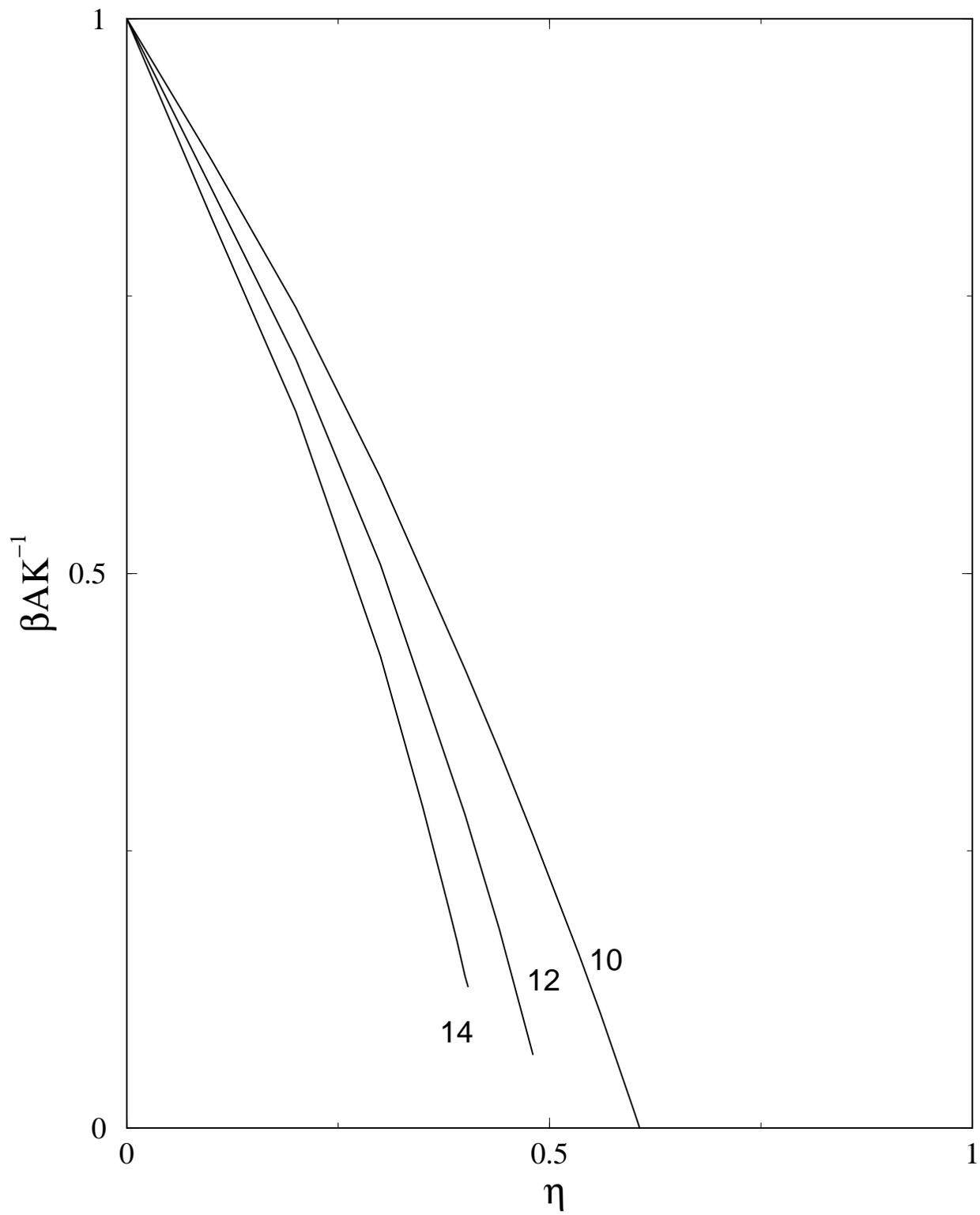
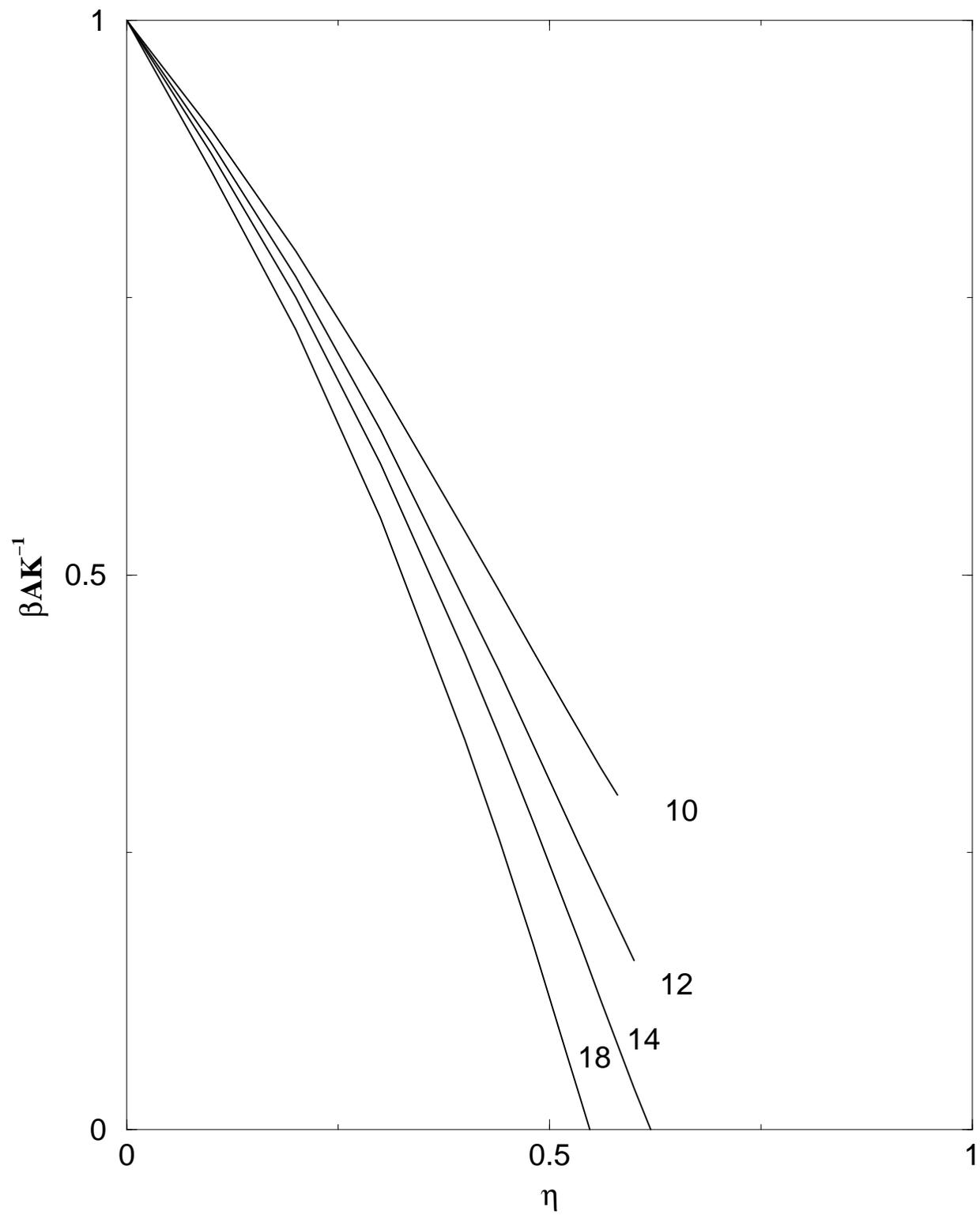
\end{document}